\documentclass[final,5p,times,twocolumn]{elsarticle}

\usepackage{amsmath}
\usepackage{amssymb}
\usepackage{xcolor}
\usepackage{graphicx}
\usepackage{float}
\usepackage{natbib}
\usepackage[colorlinks=true,linkcolor=red,citecolor=blue,urlcolor=blue]{hyperref}
\usepackage{cleveref}

\usepackage{float}

\newcommand{\yn}{\textcolor{black}}
 \newcommand{\mg}{\textcolor{black}}
\newcommand{\ms}{\textcolor{black}}
\newcommand{\nn}{\nonumber}

\begin{document}

\begin{frontmatter}

\title{Magnetars magnetic field evolution with ambipolar diffusion}

\author[iitj]{Monika Sinha}

\author[vt]{Manoj Kumar Ghosh}
\author[iitj]{Yogeesh N}

\affiliation[iitj]{
    organization={Department of Physics, Indian Institute of Technology Jodhpur},
    city={Jodhpur},
    state={Rajasthan},
    country={India}
}

\affiliation[vt]{
    organization={Department of Physics, Virginia Tech},
    city={Blacksburg},
    state={Virginia},
    country={USA}
}


\begin{abstract}
The magnetic field associated with neutron stars is
generally believed to be threaded inside the star. In the presence of a magnetic field, plasma in the interior of the star undergoes a series of processes that result in magnetic field evolution. It is believed that magnetar activities like X-ray, gamma-ray bursts, and giant flares (which release more energy in a few sec than Sun in 10 years), are mainly due to field decay. The most important process of field decay inside the core of the star is the ambipolar diffusion of the charged particles present in the interior plasma. The decay rate due to ambipolar diffusion is directly connected to the ambipolar velocity of the charged particles under the influence of the present magnetic field. The ambipolar velocity of the charged particles depends on the internal dynamics of the particles. We outline a general method to solve the particle dynamics in the presence of a magnetic field to have a magnetohydrodynamics equation from ambipolar velocity. The equation is general and applicable to all possible surrounding conditions temperature, matter states like normal or superfluid, etc.
\end{abstract}


\end{frontmatter}


\section{Introduction}
\label{level1}
\begin{figure}{}
    \centering
    \includegraphics[width=1.0\linewidth]{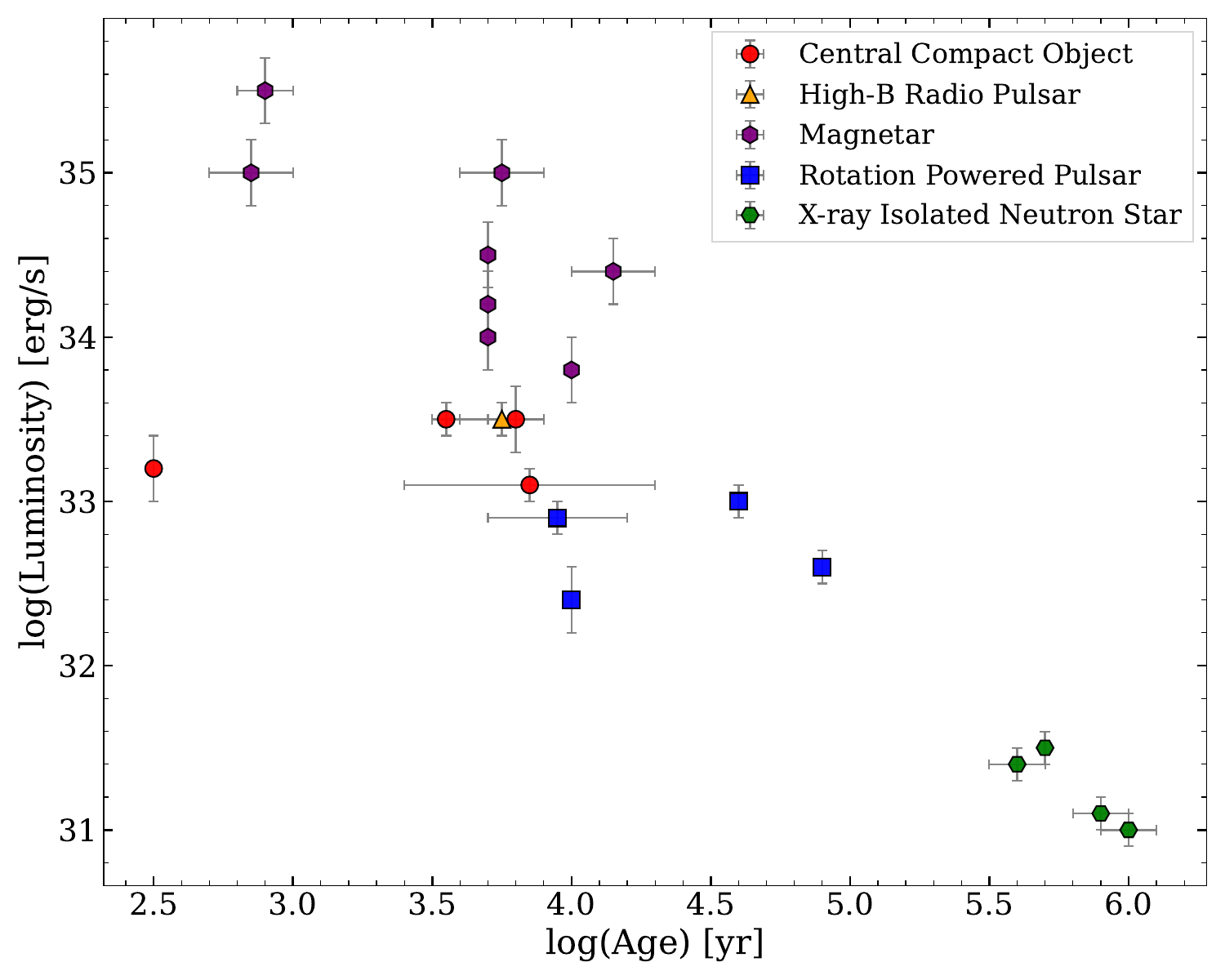}
    \caption{The plot shows the observed luminosities of \mg{NSs} as a function of their estimated ages. \mg{This data is taken from \cite{Vigano_2013}}.}
    \label{fig:magnetar_plot}
\end{figure}

Neutron stars (NSs) containing extremely dense matter possess the strongest magnetic field with the range from $10^8$ to $10^{16}$ G. The NSs with the strongest surface magnetic field in the range $10^{14}-10^{16}$ G known popularly as magnetars are comparatively younger with average age $\sim~10^3$ yrs. The NSs with medium surface field strength in the range of $10^{12}$ G are of moderate age with the age of the order of $10^6$ yrs. The older NSs with the age of the order of $10^9$ yrs have a surface magnetic field in the range of $10^{8}-10^{10}$ G. Evidently, with age the surface
magnetic field strength decreases as shown in Fig. \ref{fig:magnetar_plot}. The idea of magnetic field decay is also supported by the observational evidence of reducing surface field strength with the age of the NS class. The decay in the magnetic field with high initial field strength in magnetars is thought of as the source of high persistent luminosity in the quiescent phase of magnetars which cannot be explained as accretion powered or rotation powered \citep{1996ApJ...473..322T}.

In the presence of a magnetic field, the motion of charged particles inside the star is affected. As a result, some internal processes are initiated which ultimately lead to the evolution of the magnetic field inside the star. The mechanisms of field evolutions are mainly by three processes: 1. Ohmic decay, 2. Ambipolar diffusion 
and 3. Hall drift as discussed in detail in the classical work by \cite{1992ApJ...395..250G}.  In another novel method the generic evolution of magnetic field considering the NS matter to be non-superfluid, has been discussed in \cite{2024PhRvR...6d2050V}. The 
field evolution in the crust of the NSs is well studied 
\citep{2004ApJ...609..999C,2007A&A...470..303P,2008A&A...486..255A, 
2009A&A...496..207P, 2013MNRAS.434..123V, 2016PNAS..113.3944G,
2019MNRAS.486.4130L}. Generally, it is assumed that the field threads the interior of the star. The field evolution inside the star core has not been exhaustively studied like decay in crust.  Inside the star core, the constituent 
species are mostly neutrons with some fractions of other baryons and leptons.  The matter is in charge neutral and $\beta$-equilibrated conditions.  Depending on the temperature and field strength each of the processes dominates the field evolution inside the core. The evolution by Hall drift is dominating in stars with 
intermediate magnetic field $\sim~10^{12}-10^{13}$ G. Due to the presence of magnetic field charged particles will be in motion causing electric current and ambipolar diffusion. In this way, magnetic energy dissipation occurs due to conduction and ambipolar diffusion. Decay by ambipolar diffusion is dominating for magnetars with field strength $\sim 10^{15}$ G \citep{1998ApJ...506L..61H} 
and ohmic dissipation is dominating for low magnetic field stars with field strength $\lesssim 10^{11}$ G. Ohmic dissipation is substantial in the crust because of its low conductivity, but in the core because of high conductivity ohmic decay is not very important as far as energy dissipation is concerned.  In the core dominating decay process is ambipolar diffusion. 
Ambipolar diffusion velocity is controlled predominantly by the pressure gradient force and the collisions between different species present. 

Ambipolar diffusion plays a crucial role in various astrophysical environments beyond NSs, influencing the dynamics of magnetized plasmas on vastly different scales. In molecular clouds and star-forming regions, it regulates the gravitational collapse of gas by allowing neutral particles to drift past magnetic field lines, facilitating star formation over timescales of hundreds of thousands to millions of years \citep{1956MNRAS.116..503M, 1999ASIC..540..305M,2016A&A...596A..28B}. Similarly, in protoplanetary disks, ambipolar diffusion dominates in the outer, weakly ionized regions, affecting the evolution of magnetic fields and influencing planet formation \citep{2011ApJ...739...50B}. In highly magnetized white dwarfs, ambipolar diffusion may not contribute to long-term magnetic field evolution due to its effects are weaker compared to NSs due to lower densities and magnetic fields \citep{2002MNRAS.333..589C}. The process is also significant in brown dwarfs and giant exoplanets, where it influences atmospheric and magnetospheric interactions, particularly in gas giants like Jupiter \citep{2003IAUS..211...23B}. On cosmological scales, ambipolar diffusion may have played a role in the early universe, affecting the evolution of primordial magnetic fields in the intergalactic medium before the formation of the first stars \citep{2015MNRAS.451.2244C, 2021EPJC...81..394N}. In addition, in the magnetospheres of pulsars and magnetars, it contributes to the long-term decay and restructuring of magnetic fields, influencing their emission properties. Across these diverse astrophysical contexts, ambipolar diffusion inside NSs remains a fundamental mechanism shaping magnetic field evolution, demonstrating its broad relevance.

Ambipolar diffusion mechanism inside the core 
of the NSs has been discussed in several literature \citep{2011MNRAS.410..805G,  2011MNRAS.413.2021G,
2015MNRAS.453..671G, 2016ApJ...833..261B, 2017MNRAS.465.3416P,2017MNRAS.469.4979P,
2017MNRAS.471..507C, 2017PhRvD..96j3012G, 2018MNRAS.473.4272K, 
2018PhRvD..98d3007O, 2019MNRAS.490.2013C, 2020MNRAS.498.3000C, 2020MNRAS.499.4561G,2023MNRAS.518..821I}. Ref. 
 \cite{2011MNRAS.413.2021G} discussed the 
detailed ambipolar diffusion mechanism in both normal and superfluid matter with a comparative study of diffusion time. As further steps, others have discussed the ambipolar diffusion time scale with different choices of interactions in the superfluid matter and obtained the diffusion time. However, every calculation is limited to 
order calculation of ambipolar diffusion velocity considering some special situation that allows certain approximations. For example, \cite{2017MNRAS.471..507C} considers that the drift of the charged particles does not affect the chemical equilibrium in the interior of the star, instead, they take the variation of particle chemical potential inside the star as it is in 
the equilibrium condition. On the other hand, \cite{2017PhRvD..96j3012G} 
argued that chemical imbalance is not caused by the $\beta$-equilibrium reactions, but rather controlled by magnetic field profile only. 
In some literature \citep{2018PhRvD..98d3007O, 2020MNRAS.498.3000C} 
the problem has been treated as a perturbation in the magnetic field. With this approach, the ambipolar velocity has been estimated only from the baryon background velocity.  Moreover, the background baryon velocity determination requires the information of magnetic 
field decay which is not possible to know a priory. In another work by \cite{2020MNRAS.498.3000C} the field 
decay has been calculated neglecting the transfusion process in the low-temperature limit and considering the matter as free Fermi gas. Many literature \citep{2015MNRAS.453..671G,2017MNRAS.469.4979P,
2020MNRAS.499.4561G} discuss the ambipolar diffusion in superfluid matter discussing the nature of the force of interaction between superfluid constituents extensively. The field evolution has been simulated based on some order estimate of ambipolar diffusion \cite{2010MNRAS.408.1730H,2021CoPhC.26508001V,2023ApJ...945..151T, 2024MNRAS.528.5178S}. Ref.  \cite{2023ApJ...945..151T} used the calculation based on \cite{2016ApJ...833..261B} which assumes friction dominated case and with some order parameter. Ref. \citep{2023ApJ...945..151T, 2024MNRAS.528.5178S} discussed the ambipolar velocity but not mentioning about the superfluidity of the matter. However, the ambipolar velocity has not been determined precisely in any of them.

In the present work, we discuss a general procedure for 
solving the equations to get the ambipolar diffusion velocity profile inside the core of the star. The procedure is independent of any particular regime of temperature and is free of many approximations which have been assumed in many previous studies.
The approximation of equality of all particle velocities 
\citep{2017PhRvD..96j3012G} is not required to consider in the present calculation. 

\section{Magnetic energy and field decay}
The simplest model of NS is with the star core composed of nuclear 
matter with some fraction of electrons to maintain the charge neutrality. For massive stars, other
baryons or other phases of matter may appear in the innermost 
core of the stars. The procedure described here can be extended
for that kind of composition in straightforward manner. In general, the matter inside the NS contains mostly charge-neutral particles with some positively and negatively charged particles. So in the presence of a magnetic field, the charged particles acquire some motion which results in the ambipolar velocity of the charged particles. The part of the electric field in equilibrium which arises from the cross-product of the ambipolar velocity and the present magnetic field is responsible for the ambipolar decay of the magnetic field. The matter inside the core may be in the normal phase or in the superfluid
phase, depending on the temperature and the strength of the present 
magnetic field. 
\subsection{Normal nucleonic matter}
In normal matter the magnetic energy density decay rate is
\begin{eqnarray}
	\frac{dE_b}{dt}=\frac{d}{dt}\int d^3x \frac{B^2}{2\mu_0} 
	= \frac{1}{\mu_0} \int d^3x ~ \mathbf{B}\cdot 
	\frac{\partial \mathbf{B}}{\partial t}
	\nonumber \\
        = - \frac{1}{\mu_0} \int d^3x ~ \mathbf{B}\cdot 
	(\mathbf{\nabla} \times\mathbf{E}).
	\label{magendecay}
\end{eqnarray}
Integrating the eq. (\ref{magendecay}) by parts in steady state 
we get
\begin{equation}
\frac{dE_b}{dt} = -\frac{1}{\mu_0} \int d^3x ~ (\mathbf{\nabla}
	\times\mathbf{B})\cdot \mathbf{E}  
	= - \int d^3x ~ \mathbf{j}\cdot \mathbf{E}
 \label{magendecay1}
\end{equation}

Hence the magnetic energy density decay rate as calculated in eq. \eqref{magendecay1} due to ambipolar diffusion is 

\begin{equation}
    \left(\frac{dE_b}{dt}\right)_{amb} = - \int d^3x ~ \mathbf{j}
	\cdot(-\mathbf{V_I}\times\mathbf{B}) 
	= - \int d^3x ~ \mathbf{V}_I\cdot(\mathbf{j}\times\mathbf{B}).
 \label{rate_normal_magnetic_enr_decay}
\end{equation}
 where $\mathbf{V}_I$ is the ambipolar velocity of the charged particles. The exact expression of the ambipolar velocity $\mathbf{V}_I$ has been evaluated after discussing the proper form of interaction force in section \ref{mc}.
Then the magnetic field evolution rate due to ambipolar diffusion
is
\begin{equation}
    \left(\frac{\partial \mathbf{B}}{\partial t}\right)_{amb}
	= - \mathbf{\nabla} \times \mathbf{E}_{ambip}
	= \mathbf{\nabla}
	\times \left(\mathbf{V_I} \times \mathbf{B}\right).
    \label{field_evolution}
\end{equation}

\subsection{Superfluid nucleonic matter}
In superfluid matter magnetic energy decay rate is \citep{2015MNRAS.453..671G}
\begin{align}
\begin{split}
    \frac{dE_b}{dt} &= \frac{d}{dt}\int d^3x\frac{2\left(\mathbf{H}_{c1}\cdot \mathbf{B}\right)}{\mu_0}\\
&=\frac{2}{\mu_0}\int d^3x\left[ \mathbf{B}\cdot\frac{\partial \mathbf{H}_{c1}}{\partial t} - (\mathbf{\nabla}\times \mathbf{H}_{c1})\cdot \mathbf{E}\right]
    \label{supermagndecay}
    \end{split}
\end{align}
in the steady state, with $H_{c1}$ the lower critical field of the proton superconductivity.

Then after some algebra, (Given in appendix B) we have
\begin{align}
    \left(\frac{dE_b}{dt}\right)_{ambip} =& -  \frac{2}{\mu_0}\int Cd^3x \left\{\mathbf{V}_I \cdot \left[\left(\mathbf{\nabla}\times\mathbf{H}_{c1}\right) \times \mathbf{B}\right]\right\}
    \label{rate_super_magnetic_ener_decay}
\end{align}
where $C = \left[1 - \frac{a_{pn}\rho_p}{m_N\left(1 + \frac{\mathcal{R}_{pn}D_{pn}}{\mathcal{R}_{en}D_{en}}\right)}\right]$. 
It is apparent that the presence of proton superconductivity is affecting the ambipolar energy dissipation. 

However, neutron superfluidity doesn't affect the ambipolar energy emission rate. This energy emission has the exact same form as eq. \eqref{rate_normal_magnetic_enr_decay}.  Hence, to find out the magnetic field decay  in both cases of normal matter or superfluid matter, the main task is to find the ambipolar velocity within the star. Hence, the exact evaluation of ambipolar velocity is very important and necessary in the estimation of field decay and the corresponding time scale.

\section{General Features}
\label{level2}

We consider that the model star with a core composed of neutrons 
($n$), protons ($p$) and electrons ($e$) is rotating with angular
speed $\Omega$. Following the pioneering work on field decay mechanisms 
\cite{1992ApJ...395..250G}, we assume the Newtonian gravitational 
potential $\phi$ inside the star. The equations of motions for different species present in the star core in the presence of electromagnetic field $\mathbf{E}$ and $\mathbf{B}$, in rotating frame are
\begin{align}
	\rho_n\frac{d\mathbf{v}_n}{dt} =\rho_n\mathbf{\nabla}U -n_n \mathbf{\nabla}
	\mu_n + \mathbf{f}^{int}_n
	- 2\rho_n\mathbf{\Omega}\times\mathbf{v}_n
\label{streulnn} \\
\rho_p\frac{d\mathbf{v}_p}{dt}  = \rho_p\mathbf{\nabla}U - n_c \mathbf{\nabla} \mu_p
	+ n_c e (\mathbf{E}+\mathbf{v}_p\times\mathbf{B}) + \mathbf{f}^{int}_p \nonumber \\
	- 2\rho_p\mathbf{\Omega}\times\mathbf{v_p}
\label{streulpp} \\
\rho_e\frac{d\mathbf{v}_e}{dt} = \rho_e \mathbf{\nabla}U - n_c \mathbf{\nabla}
	\mu_e - n_c e (\mathbf{E}+\mathbf{v_e}\times\mathbf{B}) + \mathbf{f}^{int}_e \nonumber \\
	- 2\rho_e\mathbf{\Omega}\times\mathbf{v}_e
	\label{streulee}
\end{align}
where $\rho_i$ is the mass density, $\mathbf{v}_i$ the velocity in the rotating frame of the star,  $\mu_i$ the chemical 
potential, $n_i$ the number density of the particles of the $i$-th species with $n_p=n_e=n_c$ due to charge neutrality,
\begin{equation}
	U = \left(\frac{r^2\Omega^2}{2}+\phi\right)
\end{equation}
is the effective gravitational potential including the effect of
rotation and $\mathbf{f}^{int}_i$ is the mutual interaction force density on the particle of $i$-th species due to coupling with other particles.
 
We consider in general three cases, (1) both neutron and proton are in non-superfluid state when the star is sufficiently hot at the initial stage of their life, core temperature being higher than the critical temperature of superfluidity for both neutron and proton species, (2) both of them are in a superfluid state when the star cools down below the critical temperature of both the neutron and proton superfluidity, with the present magnetic field not strong enough to destroy the superconductivity, and (3) neutrons are in the superfluid state while protons are in normal state even after the core temperature goes down beyond the critical temperature of nucleon superfluidity due to the presence of strong magnetic field which is higher than the upper critical field ($H_{c2}$). Inside the NS core, the electrons are in normal state. The nature of the mutual interaction force depends on the state of matter whether it is in  superfluid or in normal phase. When the temperature falls below the critical temperature for the superfluidity of neutrons and protons, the nucleon becomes superfluid \citep{Takatsuka_1971,Anderson_1975,Jones2006,Sedrakian:2018ydt}.

Hence, inside the core, vortices of the superfluid matter are present and they contribute to the  interaction force in the form of vortex energy density and can be written as $\mathbf{f}^{int} = \mathbf{f}^c + \mathbf{f}^v$,
where $\mathbf{f}^v$ is the interaction force part due to vortex energy density and neutron vortices and
the mutual coupling force $\mathbf{f}^c$ consists of two parts - (1) mechanical collision and (2) mutual interaction with superfluid condensate and can be expressed in general as
\begin{equation}
	\mathbf{f}^c_{ij} = \mathbf{f}^{col}_{ij} + \mathbf{f}^{sf}_{ij}.
 \label{collision_force}
\end{equation}
Here$f^{sf}_{ij}$ is the mutual interaction of particles of $j$-th species with condensate of $i$-th species.  In case of normal matter $\mathbf{f}^v$ and $\mathbf{f}^{sf}_{ij}$ are zero. 

Then, in general, from eqs. \eqref{streulnn}, \eqref{streulpp} and \eqref{streulee}, in equilibrium
\begin{align}
0 &=\rho_n\mathbf{\nabla}U - n_n\mathbf{\nabla} \mu_n + \mathbf{f}^c_n
	- 2\rho_n\mathbf{\Omega}\times\mathbf{v}_n + \mathbf{f}^v_n
\label{streulnden4}\\
0 &= \rho_p\mathbf{\nabla}U - n_c\mathbf{\nabla} \mu_p + n_ce (E+\mathbf{v}_p
	\times\mathbf{B})
 + \mathbf{f}^c_p
 - 2 \rho_p\mathbf{\Omega}
	\times\mathbf{v}_p
 + \mathbf{f}^v_p
\label{streulpden4}\\
0 &= \rho_e\mathbf{\nabla}U - n_c\mathbf{\nabla} \mu_e - n_ce (E+\mathbf{v}_e
	\times\mathbf{B}) +  \mathbf{f}^c_e
 - 2\rho_e\mathbf{\Omega}
	\times\mathbf{v}_e
\label{streuleden4}
\end{align}
We have assumed that nucleons have the same mass as $m_n=m_p=m_N$. 
Adding the eq. (\ref{streulpden4}) 
and (\ref{streuleden4}) for the charged species we get

\begin{align}
\begin{split}
 0 = (\rho_N+\rho_e)n_c\mathbf{\nabla}U - n_c(\mathbf{\nabla} \mu_p
	+\mathbf{\nabla} \mu_e)
	+\mathbf{j}\times\mathbf{B} 
	+ \mathbf{f}^c_p + \mathbf{f}^c_e\\
	- 2\mathbf{\Omega}\times (\rho_p\mathbf{v}_p+\rho_e\mathbf{v}_e)
	+ \mathbf{f}^v_p
\label{strsumpeden1}
\end{split}
\end{align}
where
\begin{equation}
    \mathbf{j} = \mathbf{j}_p + \mathbf{j}_e = n_c e (\mathbf{v}_p-\mathbf{v}_e) 
    \label{ncurden1}
\end{equation}
is the total current density. With the Lorentz force density
\begin{equation}
    \mathbf{j}\times\mathbf{B}=\mathbf{f}^L
\end{equation}
the eq. (\ref{streulnden4}) and (\ref{strsumpeden1}) are

\begin{align}
  0 &= \rho_N\mathbf{\nabla}U ~-~ n_n\mathbf{\nabla} \mu_n + \mathbf{f}^c_n
 ~-~\rho_N2\mathbf{\Omega}\times\mathbf{v}_n + \mathbf{f}^v_n
\label{streulnden5} \\
  0 &= (\rho_N+\rho_e)n_c\mathbf{\nabla}U ~-~ n_c(\mathbf{\nabla} \mu_p
	+\mathbf{\nabla} \mu_e) + \mathbf{f}^L + \mathbf{f}^c_p 
	+ \mathbf{f}^c_e \notag \\
    &\quad -~ 2\mathbf{\Omega}\times (\rho_p\mathbf{v}_p+\rho_e\mathbf{v}_e)
+ \mathbf{f}^v_p
\label{peq2}
\end{align}

\subsection{Coupling force density}
Let us examine the coupling force density. Suppose force on 
one $i$-th particle due to mutual coupling with one $j$-th particle 
is $\mathbf{F}^c_{ij}$. 
Now the mutual coupling force on $i$-th particle due to its interaction with other particles is $\mathbf{F}^c_i=\Sigma_j\mathbf{F}^c_{ij}$.
Within unit volume the force density for $i$-th particle is $\mathbf{f}^c_i=n_i\Sigma_j\mathbf{F}^c_{ij}$.
Then
\begin{equation}
\mathbf{f}^c_n=  n_n\mathbf{F}^c_n = n_n\left(\mathbf{F}^c_{np}+\mathbf{F}^c_{ne}\right)
\label{nfden}
\end{equation}
\begin{eqnarray}
{\rm and}~~~~~~ \mathbf{f}^c_p=n_p\left(\mathbf{F}^c_{pn}+\mathbf{F}^c_{pe} \right)
 \label{pfricden} \\
\mathbf{f}^c_e=n_e\left(\mathbf{F}^c_{ep}+\mathbf{F}^c_{en}\right)
\nonumber \\
=-n_e\mathbf{F}^c_{pe}+n_e\mathbf{F}^c_{en}
\label{efricden} 
\end{eqnarray}
as according to Newton's third law $\mathbf{F}^c_{ij}=-\mathbf{F}^c_{ji}$.
Hence, 
\begin{equation}
\mathbf{f}^c_p+\mathbf{f}^c_e=n_c\mathbf{F}^c_{pn}+n_c\mathbf{F}^c_{en}= -n_c\mathbf{F}^c_n
\label{cfricden}
\end{equation}

Hence,
\begin{equation}
    -\frac{\mathbf{f}_p^c + \mathbf{f}_e^c}{n_c} = \frac{\mathbf{f}_n^c}{n_n} =  \mathbf{F}^c_n
\end{equation}
Now dividing the eq. \eqref{streulnden5} by neutron mass density 
$\rho_n=m_Nn_n$ and the eq. \eqref{peq2} by proton mass 
density $\rho_p=m_Nn_c$ we have

\begin{eqnarray}
0=\mathbf{\nabla}U ~-~ \mathbf{\nabla}\left(\frac{\mu_n}{m_N}\right)
	+\frac{\mathbf{F}^c_n}{m_N} ~+~ \frac{\mathbf{f}^v_n}{\rho_n} ~-~ 2\mathbf{\Omega}\times\mathbf{v}_n
   \label{strnacc}
\\
    0 = \mathbf{\nabla}U ~-~\mathbf{\nabla}\left(\frac{\mu_p+\mu_e}{m_N}\right)
	+\frac{\mathbf{f}^L}{\rho_p}-\frac{\mathbf{F}^c_n}{m_N} ~ +~ \frac{\mathbf{f}^v_p}{\rho_p}
	-~ 2\mathbf{\Omega}\times \mathbf{v}_p
    \label{strsumpeacc}
\end{eqnarray}
as $m_e\ll m_N$.

Then subtracting the eq. (\ref{strsumpeacc}) from the eq. 
(\ref{strnacc}) and multiplying by proton fraction $x_p=\rho_p/\rho$ 
we get

\begin{equation}
	0 =x_p\mathbf{\nabla}\beta ~+~ 2 x_p \frac{n_n\mathbf{F}_n^c}{\rho_n} ~
	+ \frac{\rho_p\mathbf{f}^v_n}{\rho_n \rho}~-~ \frac{\mathbf{f}^v_p}{\rho}+~ 2x_p\mathbf{\Omega}\times \mathbf{V}_p  - \frac{\mathbf{f}^L}{\rho} \label {strsumpedeffnacc}
\end{equation}
where $\beta=\left(\mu_p+\mu_e-\mu_n\right)/m_N$ is the chemical 
imbalance per unit mass and $\mathbf{V}_p=\mathbf{v}_p-\mathbf{v}_n$ 
is the proton velocity in the neutron rest frame. Therefore, in addition 
to the rotation, the Lorentz force is balanced by force due to (1) 
Transfusion, (2) Mutual coupling, and if nucleon vortices are 
present by (3) Vortex tension. Now the transfusion term appeared as the first term of eq.  
(\ref{strsumpedeffnacc}) is related to the weak interaction maintaining 
the $\beta$-equilibrium. The coupling force appearing in the 
second term of the same equation as well as the transfusion 
terms are related to the participating particle velocity. The 
vortex tension is present only for superfluid matter, They are 
discussed in the subsequent sections.

\section{Transfusion}

The chemical equilibrium is disturbed due to the motion of charged 
particles in the presence of the magnetic field. That creates a gradient 
of imbalanced chemical potential which acts to balance the magnetic 
force on the charged particles. Chemical imbalance starts the 
weak interaction $n~\longleftrightarrow~p~+~e~+{\bar\nu_e}$.
If the reaction rate is $\delta\Gamma=\lambda\beta=\Gamma(p+e^
-\to n+\nu_e) -\Gamma(n \to p+e^-+{\bar\nu_e})$ then the charged 
particles follow the equation of continuity
\begin{equation}
    \frac{\partial n_c}{\partial t} + n_c\mathbf{\nabla}\cdot 
	\mathbf{v}_c = -\lambda\beta.
\end{equation}
The above equation is specific to matter with uniform density. 
hence at each point inside the star, the equation is valid. Then 
in steady state \citep{1992ApJ...395..250G} $n_c\mathbf{\nabla}\cdot\mathbf{v}_c = -\lambda\beta$
for both the charged species. Then adding these equations for 
protons and electrons we get

\begin{equation}
    n_c\mathbf{\nabla}\cdot \mathbf{w}
= -\lambda\beta 
\label{transfusion}
\end{equation}
where
\begin{equation}
    \mathbf{w} = \frac{\mathbf{v}_p+\mathbf{v}_e}{2}.
    \label{avchvel}
\end{equation}

\ms{For superfluid nucleons the reaction rate is reduces by a factor $R_{sf}$ as}
\mg{\begin{align}
     n_c\mathbf{\nabla}\cdot\mathbf{w} = -R_{sf}\lambda\beta
\end{align}}
\yn{where $R_{sf} \approx 3 \times 10^-4\,y_{p}^5\,e^{-2y_n}$ \citep{2011MNRAS.413.2021G} with $y_p = 1.764\,T_{cp}/T$, $y_n = 1.764\,T_{cn}/T$, where $T_{cp}$ and $T_{cn}$ are the critical temperatures of proton and neutron superfluidity respectively.}

\section{ Mutual couplings}
\subsection{Normal nucleonic matter\label{sec: normal_matter}}
At the younger age of NS, as the core temperature is high, the nucleons are in non-superfluid phase.  In this case, $\mathbf{f}^c$ is only due to mechanical collision
    $\mathbf{f}^c = \mathbf{f}^{col}_{ij} = - D_{ij} \left(\mathbf{v}_i 
	- \mathbf{v}_j\right)$
where the $f^{col}_{ij}$ is the force due to mechanical collision between the particles of $i$-th and $j$-th species and $D_{ij}$ is the collision coefficient.

\subsection{\label{sec: superfluid_nucleonic_matter}Superfluid nucleonic matter}
In this case as mentioned by eq. \eqref{collision_force} the mutual coupling terms are combination of mechanical collision between two particles and interaction with condensates.

The mechanical collision force in this case can be expressed as proportional to their relative velocity as in normal matter case, however, reduced due to the superfluid nature of matter by a factor as \citep{2011MNRAS.413.2021G}
\begin{equation}
	\mathbf{f}^{col}_{ij} = - {\cal R}_{ij} D_{ij} \left(\mathbf{v}_i 
	- \mathbf{v}_j\right)
  \label{coupling_force}
\end{equation}
where ${\cal R}_{ij}$ is the reduction factor due to the presence of superfluidity. 
The force on particle of $i$-th species due to interaction with $\alpha$-th nucleon condensate \citep{1995apj...447..305s} is
\begin{align}
	\mathbf{f}^{sf}_{\alpha i} = a_{\alpha i}\left(\mathbf{\omega^0_\alpha}
	\times\mathbf{p}_{\alpha i}\right)~ +~b_{\alpha i}
	\left\{\hat{\omega}_\alpha \times\left(\mathbf{\omega^0_\alpha} 
	\times\mathbf{p}_{\alpha i}\right)\right\}
	\nonumber \\
	-~c_{\alpha i}
	\left(\mathbf{\omega^0_\alpha} \cdot\mathbf{p}_{\alpha i}\right)
\label{forcesuperfluid}
\end{align}
Here the index $\alpha$ indicates the species of condensate i.e. $\alpha=N=n,p$ and 
\begin{eqnarray}
\mathbf{\omega^0_\alpha} = \mathbf{\omega_\alpha} + \frac{e_\alpha}{m_nc}\mathbf{B}
\label{notation1},\\
	\mathbf{p}_{\alpha i} = \rho_\alpha(\mathbf{v}_\alpha-\mathbf{v}_i)
	+~\mathbf{\nabla}\times\lambda_\alpha\mathbf{\hat{\omega}}_\alpha.
    \label{noennotation3} 
\end{eqnarray}
in the absence of entrainment
with $\mathbf{\omega_N}$ the circulation quanta of the superfluid nucleons. Here $e_N\equiv (0,e)$ is the electric charge of neutrons and protons and $\lambda_N$ is the angular momentum of the circulation. The contribution from the first term of the eq. (\ref{forcesuperfluid}) is mainly coming due to the interaction between flux tube and vortices, and  the second term comes mainly due to the interaction of normal charged particles scattering off the vortices. Since the component of particle velocity along the vortex line is small, we shall neglect the third term of the equation (\ref{forcesuperfluid})\citep{1995apj...447..305s}.
Moreover for neutrons $\mathbf{\omega}^0_n = \mathbf{\omega}_n$.

The non-existence of proton superconductor due to an intense magnetic field gives rise to the third case. In this situation,  $\mathbf{f}^c$ is similar to the eq. \eqref{collision_force} and so is collision force eq. \eqref{coupling_force}. However, for $\mathbf{f}^{sf}$, only 2nd term in eq. \eqref{forcesuperfluid} will be present due to the absence of proton superconductivity. So, force is
\begin{equation}
    \mathbf{f}^{sf}_{\alpha i} = ~b_{\alpha i}
	\left\{\hat{\omega}_\alpha \times\left(\mathbf{\omega^0_\alpha} 
	\times\mathbf{p}_{\alpha i}\right)\right\}
\end{equation}
and the rest of the forces is exactly similar to the superfluid nucleonic matter.

\subsection{Putting together}\label{mc}
The mutual coupling force as given  Eq. \eqref{nfden} can be calculated from Eqs. \eqref{collision_force}, \eqref{coupling_force} and \eqref{forcesuperfluid}

\begin{align}
\begin{split}
\mathbf{f}_n^c
	&= {\cal R}_{np}D_{np}\mathbf{V}_p + {\cal R}_{ne}D_{ne}\mathbf{V}_e  \\
	&~+~  \mathbf{\omega}_n\times(a_{np}\mathbf{p}_{n p} 
	+a_{ne}\mathbf{p}_{n e})\\
	&~+~\hat{\omega}_n\times\{ \mathbf{\omega}_n\times(b_{np}
	\mathbf{p}_{n p} +b_{ne} \mathbf{p}_{n e})\}.
	\label{fc}
 \end{split}
\end{align}

Now
\begin{align}
\begin{split}
\mathbf{\omega_n}\times(a_{np}\mathbf{p}_{n p} 
	+a_{ne}\mathbf{p}_{n e}) 
=\mathbf{\omega_n} \times (a_{np}+a_{ne})\rho_n\mathbf{v}_n \\
	~-~ \mathbf{\omega}_n \times \rho_n(a_{np}\mathbf{v}_p
	+a_{ne}\mathbf{v}_e)
	\label{cross1}
 \end{split}
\end{align}
and
\begin{align}
\begin{split}
&\hat{\omega}_n\times\{ \mathbf{\omega}_n\times(b_{np}\mathbf{p}_{n p} 
	+b_{ne}\mathbf{p}_{n e})\}\\
	= &\hat{\omega}_n\times\left[\mathbf{\omega}_n\rho_n \times 
	\{(b_{np}+ b_{ne})\mathbf{v}_n ~\right.
	\left.-~
(b_{np}\mathbf{v}_p+ b_{ne}\mathbf{v}_e)\}\right]
	\label{2cross}
 \end{split}
\end{align}
noting
\begin{align}
\begin{split}
a_{np}\mathbf{p}_{n p} + a_{ne}\mathbf{p}_{n e}
= &\rho_n(a_{np}+ a_{ne})\mathbf{v}_n - \rho_n(a_{np}\mathbf{v}_p
	+a_{ne}\mathbf{v}_e) \\
	&~+~ (a_{np}+ a_{ne}) \mathbf{\nabla}\times\lambda_n\mathbf{\hat{\omega}_n},
 \end{split}
\end{align}
from eq. \eqref{noennotation3}.
Then in the rest frame of neutrons from eq. (\ref{cross1})
\begin{align}
\begin{split}
	\mathbf{\omega}_n\times\left(a_{np}\mathbf{p}_{n p} 
	+a_{ne}\mathbf{p}_{n e}\right) &= 
	-~ \mathbf{\omega}_n\times \rho_n(a_{np}\mathbf{V}_p
	+ a_{ne}\mathbf{V}_e)\\
 & =  -~
\mathbf{\omega}_n \times \rho_n a\mathbf{V}_1
\end{split}
\end{align}
and from eq. (\ref{2cross})
\begin{align}
\begin{split}
&\hat{\omega}_n\times\{ \mathbf{\omega}_n\times(b_{np}\mathbf{p}_{n p} 
	+ b_{ne}\mathbf{p}_{n e})\} \\
	&= \hat{\omega}_n\times [\mathbf{\omega}_n\rho_n\times\{-~
(b_{np}\mathbf{V}_p+ b_{ne}\mathbf{V}_e)\}]\\
 &= \hat{\omega}_n\times [\mathbf{\omega}_n\rho_n\times\{-~
    b\mathbf{V}_2\}].
\end{split}
\end{align}
Hence in the rest frame of neutrons from eq. (\ref{fc})
\begin{equation}
    \mathbf{f}^c_n = D_I\mathbf{V}_I ~ -~
\mathbf{\omega}_n \times \rho_n a\mathbf{V}_1 ~-~ \hat{\omega}_n\times [\mathbf{\omega}_n\rho_n\times b\mathbf{V}_2]
\label{finalfc}
\end{equation}
where we define
\begin{align}
{\cal R}_{np}D_{np}\mathbf{V}_p + {\cal R}_{ne}D_{ne}\mathbf{V}_e = D_I\mathbf{V}_I,
\label{ambiii} \\
a_{np}\mathbf{V}_p+ a_{ne}\mathbf{V}_e = a\mathbf{V}_1,
\\
b_{np}\mathbf{V}_p+ b_{ne}\mathbf{V}_e = b\mathbf{V}_2.
\label{ambi vel 3}
\end{align}
with
\begin{eqnarray}
    D_I = {\cal R}_{np}D_{np} + {\cal R}_{ne}D_{ne}, 
    \label{DI} \\
    a = a_{np} + a_{ne}, 
    \\ 
    {\rm and} ~~~ b = b_{np} + b_{ne}
\end{eqnarray}

Then finally we can write the eq. (\ref{strsumpedeffnacc}) as

\begin{align}
\begin{split}
0 &= x_p\mathbf{\nabla}\beta + \frac{2x_p}{\rho_n} [D_I\mathbf{V}_I - \mathbf{\omega}_n \times \rho_n a\mathbf{V}_1 - \hat{\omega}_n\times (\mathbf{\omega}_n\rho_n\times b\mathbf{V}_2)]\\
&~+~ \frac{\rho_p\mathbf{f}^v_n}{\rho_n \rho}
~-~ \frac{\mathbf{f}^v_p}{\rho}+~ 2x_p\mathbf{\Omega}\times 
\mathbf{V}_p  - \frac{\mathbf{f}^L}{\rho}
\label{strsumpedeffnaccsfii}
\end{split}
\end{align}

Now inside the NS core, the electrons are in normal
phase. in this situation we can consider $a_{ne}\approx 0$.
Then we can take 

\yn{
    $V_{p} = V_{1}$ and $a_{np} = a$.}

With the total current density 
\begin{equation}
 \mathbf{j} = \mathbf{j}_p + \mathbf{j}_e = n_c e (\mathbf{v}_p-\mathbf{v}_e)
    \label{ncurden}
\end{equation}
From eq. (\ref{ambiii}) we can express proton
and electron velocity in neutron rest frame in terms of ambipolar
velocity $\mathbf{V}_I$ and current density $\mathbf{j}$ as
\begin{equation}
\mathbf{V}_p = \mathbf{V}_I + A \mathbf{j}
	~~~~~~~~~~~~~~~\text{and}
~~~~~~~~~~~~~\mathbf{V}_e = \mathbf{V}_I - B \mathbf{j}.
\label{eq:vpve}
\end{equation}
where
\begin{eqnarray}
	A = \frac{\frac{{\cal R}_{ne}D_{ne}}{en_{c}}}{D_I}~~~~~~B = \frac{\frac{{\cal R}_{np}D_{np}}{en_{c}}}{D_I}
    \label{eq:ab}
\end{eqnarray}
With this the eq. (\ref{strsumpedeffnaccsfii}) becomes

\begin{align}
\begin{split}
	0 &= x_p\mathbf{\nabla}\beta ~+~ \frac{2x_p}{\rho_n}\left(
	D_I\mathbf{V_I} ~-~ \mathbf{\omega_n} \times \rho_n a
	(\mathbf{V}_I + A \mathbf{j})\right)\\
	&~-~\frac{2x_p}{\rho_n}\left(\hat{\omega}_n\times 
	[\mathbf{\omega}_n\rho_n\times\{b_{np}(\mathbf{V}_I 
	+ A \mathbf{j}) + b_{ne} (\mathbf{V}_I 
	- B \mathbf{j})\}]\right)\\
&+ \frac{\rho_p\mathbf{f}^v_n}{\rho_n \rho}~-~ \frac{\mathbf{f}^v_p}{\rho} + x_p2\mathbf{\Omega}\times (\mathbf{V}_I + A \mathbf{j}) - \frac{\mathbf{f}^L}{\rho}
\label{strsumpedeffnaccsfgen}
\end{split}
\end{align}
In general, the equation can be solved by having all the terms 
present theoretically by numerical technique for a given magnetic 
field profile inside the star. The term $\nabla\beta$ appearing in the above equation can be
expressed in terms of proton and electron velocities and hence,
in terms of ambipolar velocity $\mathbf{V}_I$ and current density
$\mathbf{j}$. Consequently, the equation involves only ambipolar
velocity as unknown with the current density given by
\begin{equation}
	\mathbf{j} = \frac{1}{\mu_0} \mathbf{\nabla}\times \mathbf{B}
\end{equation}
for a given magnetic field profile in case of normal matter 
and in the case of superconducting fluid, the current density is
given by \citep{2011MNRAS.410..805G}
\begin{equation}
	\mathbf{j} = \frac{1}{\mu_0} \mathbf{\nabla}\times \mathbf{B}_L
\end{equation}
where, $\mathbf{B}_L$ is London field.

However, in real cases, the interior 
of the star cools down, and with stars temperature decreasing, 
different processes happen to be dominant as well as the matter 
property changes. Therefore, the matter property and the surrounding 
conditions select some of the terms of the eq. (\ref{strsumpedeffnaccsfgen}) 
naturally and discards the rest. Consequently, in a realistic 
case procedure to solve for ambipolar velocity becomes easier 
for different situations prevailing inside the core of the star. 
The special cases are discussed theoretically in the following sections.

\section{Special cases}
\subsection{Normal nucleonic matter}
As already discussed, for younger NSs when the core temperature is higher than $T_c$ both the neutrons and protons are in a normal state. Then in the case of normal matter, the nucleon vortices are absent. So the superfluid interaction force is absent. Then the collision force is simply (from eq. \eqref{finalfc})
\begin{equation}
    \mathbf{f}^c = D_I \mathbf{V}_I
\end{equation}
where from eq. \eqref{ambiii}
\begin{equation}
    D_I\mathbf{V}_I = D_{np} \mathbf{V}_p + D_{ne}\mathbf{V}_e
    \label{nambipolar_velocity}
\end{equation}
and from eq. \eqref{DI}
\begin{equation}
    D_I = D_{np} + D_{ne}
\end{equation}
As $\cal{R}$'s are $1$ for normal matter. Also as there is no
vortex tension force, the final eq. \eqref{strsumpedeffnaccsfgen} in equilibrium for normal matter is
\begin{equation}
	0 = x_p\mathbf{\nabla}\beta ~+~ \frac{2x_p}{\rho_n}
	D_I\mathbf{V_I} + 2x_p\mathbf{\Omega}\times (\mathbf{V}_I + A \mathbf{j}) - \frac{\mathbf{f}^L}{\rho}
\label{eq:normal_matter}
\end{equation}

\subsection{Supefluid nucleonic matter}
With time when the temperature of the star core goes down below $T_c$ for neutrons and protons, the matter becomes superfluid.

\textbf{Vortex tension force density:}
If we consider the nucleons are superfluid, the present 
magnetic field is quantized inside the vortices of 
the nucleons. Then the vortex tension force for the neutron
and proton vortices are \citep{2011MNRAS.410..805G}

\begin{equation}
    \mathbf{f}^v_n = \frac{\rho_n}{\mu_0 en_n}\left[ - \mathbf{\omega_n} \times \left\{\nabla\times(H_{vn}\hat{ \omega}_n) \right\} + \frac{n_n}{\rho_n} \left(\nabla\times\mathbf{p}_n\right) \times \left(\nabla \times \mathbf{B_L}\right) \right] 
\end{equation}
and
\begin{align}
\begin{split}
    \mathbf{f}^v_p = \frac{\rho_p}{\mu_0 en_c}\left[ - \mathbf{\omega_p} 
	\times \left\{\nabla\times(H_{vp}\hat{\omega}_p) \right\} 
	+ \frac{n_c}{\rho_p} \left(\nabla\times\mathbf{p_p}\right) 
	\times \left(\nabla \times \mathbf{B}_L\right) \right]\\
	- \frac{1}{\mu_0} (\nabla \times \mathbf{B}_L) \times \mathbf{B}
    \label{tp}
\end{split}
\end{align}

respectively, where $\mathbf{\omega_i}$ 
is the circulation per unit area, the circulation quanta for 
the $i$-th species and 
\begin{equation}
    H_{vi} = \frac{\mu_0\mathcal{E_{\mathrm{vi}}}}{\phi_0} 
\end{equation}
with $\phi_0 =\frac{h}{2e} $ \citep{tilleyandtilley} is the flux quanta and 
\begin{equation}
    \mathcal{E}_{vn} =   \frac{\kappa^2}{4\pi}\frac{\rho_n}{1 - \epsilon_n}log\left(\frac{d_n}{a_n}\right) \qquad
    \mathcal{E}_{vp} = \frac{\kappa^2 \rho_p}{4\pi \tilde m_p^{*}}log\left(\frac{b}{a_p}\right)
\end{equation}
with $\tilde{m_p}^* = m_p^*/m_N$ and $\kappa = \frac{h}{2m} $ \citep{tilleyandtilley} the vorticity of each vortex, $d_n$ the
inter vortex distance for neutron vortices, $a_i$ is the 
coherence length and $b$ is penetration length. For proton 
vortices, with certain approximations it can be shown that \citep{2011MNRAS.410..805G}
\begin{equation}
H_{vp} = H_{c1}
\end{equation}
where $H_{c1}$  is lower critical field for 
proton superconductivity. In the presence of superconducting
matter the total current 
\begin{equation}
    \mathbf{j} = \frac1{\mu_0} \nabla\times\mathbf{B_L}.
\end{equation}
Hence, the Lorentz force density 
\begin{equation}
    \mathbf{f}^L = \mathbf{j}\times\mathbf{B} = \frac1{\mu_0} (\nabla\times\mathbf{B_L})\times\mathbf{B}. 
\end{equation}
Here due to proton superconductivity in eq. \eqref{ambi vel 3} first term will not be there. So force due to nucleon condensate will be 
\begin{equation}
    \mathbf{f}^c = D_I\mathbf{V}_I ~ -~
\mathbf{\omega}_n \times \rho_n a\mathbf{V}_1 ~-~ \hat{\omega}_n\times [\mathbf{\omega}_n\rho_n\times b\mathbf{V}_2]
\end{equation}
where we have defined $b_{ne}\mathbf{V}_e = b\mathbf{V}_2$  and from eq. \eqref{ambiii}
\begin{equation}
    {\cal R}_{np}D_{np}\mathbf{V}_p + {\cal R}_{ne}D_{ne}\mathbf{V}_e = D_I\mathbf{V}_I
    \label{superfluid_ambipolar}
\end{equation}

Then putting the expression of $\mathbf{f}^v_p$ and $\mathbf{f}^L$ in eq. \eqref{strsumpedeffnaccsfgen} we have
for superfluid nuclear matter

\begin{align}
\begin{split}
0 &= x_p\mathbf{\nabla}\beta + \frac{2x_p}{\rho_n}\left[D_I\mathbf{V}_I - \mathbf{\omega}_n \times \rho_n a\left(\mathbf{V}_I + A\mathbf{j}\right)\right]\\
 &- \frac{1}{\rho_n}\left[\hat{\omega}_n \times \left[\mathbf{\omega}_n\rho_n \times \left\{b_{ne}\left(\mathbf{V}_I - B\mathbf{j}\right)\right\}\right]\right]\\
 &~+~ \frac{\rho_p}{n_n \rho}\frac{1}{\mu_0 e}\left[ - \mathbf{\omega_n} \times \left\{\nabla\times(H_{vn}\hat{ \omega}_n) \right\} + \frac{n_n}{\rho_n} \left(\nabla\times\mathbf{p}_n\right) \times \left(\nabla \times \mathbf{B_L}\right) \right]\\
 &~-~ \frac{\rho_p}{n_c\mu_0 e\rho}\left[~-~\mathbf{\omega}_p \times (\mathbf{\nabla}\times (H_{c1}\hat{\mathbf{\omega}}_p))\right]\\
 &~-~  \frac{\rho_p}{n_c\mu_0 e\rho}\left[\frac{n_c}{\rho_p}\left(\mathbf{\nabla} \times \mathbf{p}_p\right) \times \left( \mathbf{\nabla} \times \mathbf{B}_L\right) \right]\\
&~+~ 2x_p\mathbf{\Omega}\times (\mathbf{V}_I + A \mathbf{j}) 
\label{eq:final1}
\end{split}
\end{align}

\subsection{Superfluid neutron and normal proton matter}
Another possible scenario for magnetar is that in the presence of a strong magnetic field is that if the upper critical field of proton superconductivity is less than the present magnetic field then proton superconductivity is absent.
 If we consider only neutrons are superfluid and both protons and electrons are normal matter then tension force due to proton vortices $\mathbf{f}^v_p$ vanishes and neutron vortices are not magnetized. Then neutron vortex tension force is
\begin{equation}
    \mathbf{f}^v_n = \frac{\rho_n}{n_n\mu_0 e}\left[ - \mathbf{\omega_n} \times \left\{\nabla\times(H_{vn}\hat{ \omega}_n) \right\}\right]
\end{equation}
 and the proton fluxtubes are absent. Hence, in the expression of force due to the presence of nucleon condensate, the first term is absent. Then the force of collision as in eq. \eqref{finalfc} reduces to 
\begin{equation}
    \mathbf{f}^c = D_I\mathbf{V}_I 
- \hat{\omega}_n\times [\mathbf{\omega}_n\rho_n\times b\mathbf{V}_2].
\label{fcnp}
\end{equation}

Now putting the values of $\mathbf{f}^v_n$ and 
$\mathbf{f}^c$  in eq. \eqref{strsumpedeffnaccsfgen} we obtain the equation for only superfluid neutron matter,

\begin{align}
0 &= x_p\mathbf{\nabla}\beta + \frac{2x_p}{\rho_n}\biggl(D_I\mathbf{V_I} - \hat{\omega}_n\times [\mathbf{\omega}_n\rho_n\times\{b_{np}(\mathbf{V}_I + A \mathbf{j})\biggl)\nn\\
&-\frac{2x_p}{\rho_n}\biggl(\hat{\omega}_n\times [b_{ne} (\mathbf{V}_I - B \mathbf{j})\}]\biggr)\nn\\
&+\frac{\rho_p}{n_n \rho}\frac{1}{\mu_0 e}\biggl[- \mathbf{\omega_n} \times \biggl\{\nabla\times(H_{vn}\hat{ \omega}_n) \biggr\} \biggr]
- \frac{1}{\mu_0}\biggl(\mathbf{\nabla} \times \mathbf{B}\biggr) \times \mathbf{B}\nn\\
&+ 2x_p\mathbf{\Omega}\times (\mathbf{V}_I + A \mathbf{j}) 
\label{eq:final2}
\end{align}

\section{Realistic scenerio}

Generally, magnetars are young NSs with a surface magnetic field of the order $10^{14-16}$ G. They are slow rotators with a period of the order of $10$s. Magnetars are exceptionally hot with surface temperature $\sim 5\times 10^6$K \citep{2016ApJ...833..261B,2023ApJ...945..151T}. Such high surface temperature can be sustained with core temperature in the order of $10^{8-9}$K depending on the composition and field strength in the atmosphere \citep{2016ApJ...833..261B,2004ARA&A..42..169Y}. Inside the core of the stars, the neutrons may form Cooper pair in $^3P_2$ channel creating neutron superfluid matter and protons form Cooper pair in $^1S_0$ channel \cite{Anderson_1975,Takatsuka_1971,Sedrakian:2018ydt}. With the existing pairing interactions the critical temperature of neutron superfluidity ($T_{cn}$) inside the core of NS is in the range $10^{8-9}$K and the critical temperature of proton superconductivity ($T_{cp}$) is above $10^9$K for all permissible models which vary with the density of matter inside the core as well. On the other hand, proton superconductivity also depends on the present magnetic field strength. If the present magnetic field is higher than the upper critical field $H_{c2}$ then the proton pairing will be destroyed. Then depending upon the field profile inside the star, some the portion of the star will be free of proton superconductivity and some portion will contain proton paring. Hence, depending on the considered models, neutron superfluidity and proton superconductivity will be present inside the entire core or in part of the core.

\section{Model}

In this work we consider magnetars as made of nucleonic matter constructed with DD-ME2 equation of state (EoS) within covariant density functional model. This EoS is consistent with several astrophysical observations \citep{Demorest2010,vanKerkwijk2011,Vivek2021} including mass-radius constraints as well as constraints from gravitational wave observations. We consider two model stars with masses $1.4$ and $2.3~M_\odot$. We consider the star rotating with a period of $10$s, typical for magnetars. 
\begin{figure}
  \centering
    \includegraphics[width=\linewidth]{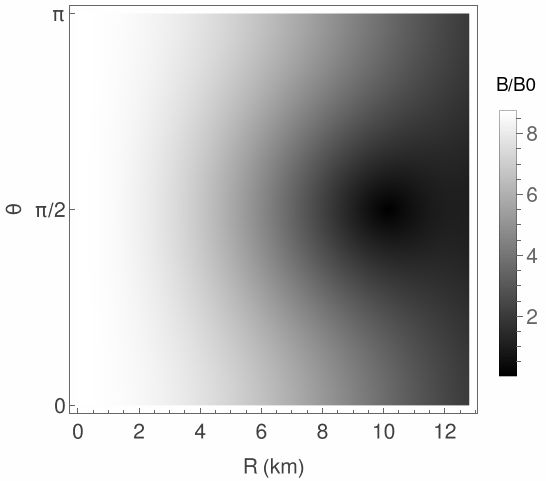}
    \caption{Magnetic field profile of the star.}
    \label{fig:1.4_magnetic_field}
\end{figure}

The magnetic field configuration we are taking is discussed below

\subsection{Magnetic field profile}
The magnetic field's precise profile inside an NS is unknown. We have taken an axisymmetric poloidal magnetic field that satisfies the relevant boundary and 
regularity condition \citep{2013MNRAS.433.2445A}

\begin{equation}
    \mathbf{B} = B_0\left[\frac{2f(x)cos\theta}{x^2}\hat{r} - \frac{f'(x)sin\theta}{x} \hat{\theta}\right]
    \label{magnetic_field_profile}
\end{equation}
where $B_0$ is the magnetic field strength at the centre of the star and $x$, is a dimensionless quantity; $x = r/R$, where $R$ is the radius of the star. The function $f(x)$ is given by
\[
f(x) = \begin{cases}
\frac{35}{8}x^2 - \frac{21}{4}x^4 + \frac{15}{8}x^6 & \text{if } x < 1 \\
x^{-1} & \text{if } x > 1.
\end{cases}
\]

The magnetic field profile inside the star is shown in fig. (\ref{fig:1.4_magnetic_field}).

Here, to have the surface field strength of the order of $10^{15}$ G, we have taken $B_0=4.42\times10^{15}$ G and $B_0=4.31\times10^{15}$ G for $1.4$ and $2.3~M_\odot$ respectively.

\subsection{Pairing models}

\begin{figure}
  \centering
    \includegraphics[width=\linewidth]{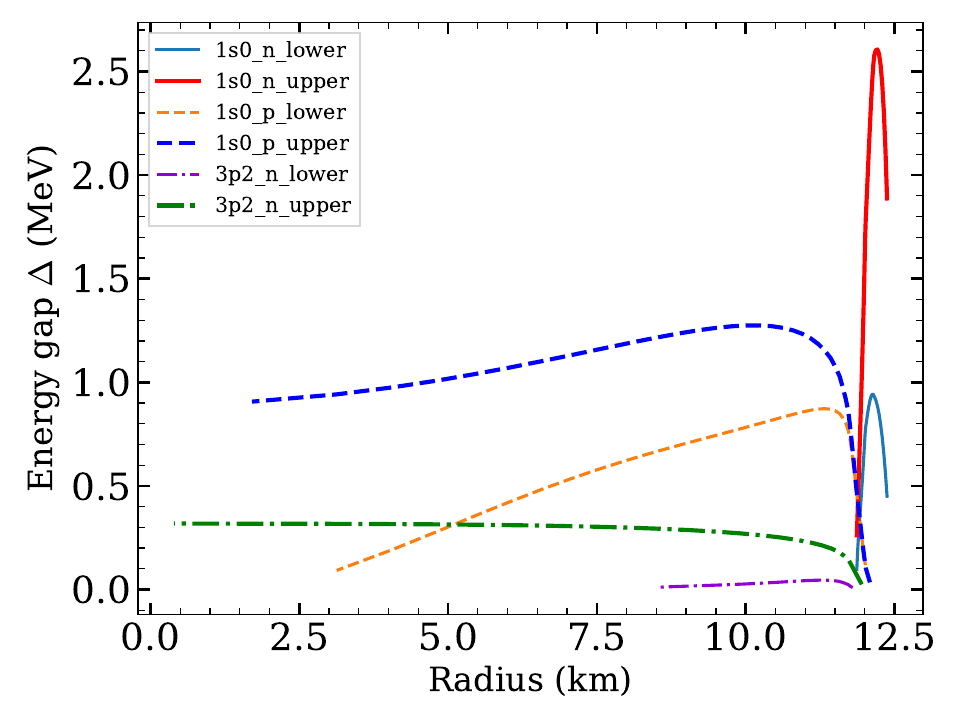}
    \caption{\yn{Superfluid pairing energy gap for 1.4$M_{\odot}$ star}}
    \label{fig:1.4_delta}
\end{figure}

\begin{figure}[t]
    \includegraphics[width=\linewidth]{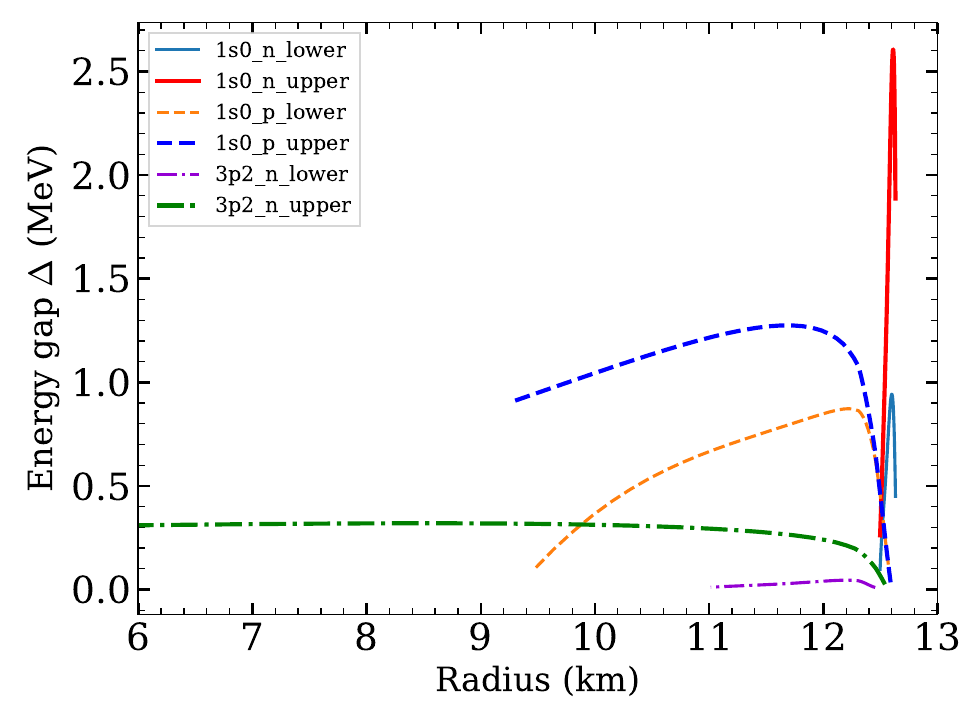}
    \caption{\yn{Superfluid pairing energy gap for 2.3$M_{\odot}$ star}.}
    \label{fig:2.3_delta}
\end{figure}

We consider the energy gap of neutrons ($\Delta_n$) and protons ($\Delta_p$) as discussed in ref. \cite{2019EPJA...55..167S}. The variation of the energy gap of neutrons ($\Delta_n$) and protons ($\Delta_p$)  with radius (R) for different pairing types in a $1.4~M_\odot$ star are shown in fig. \ref{fig:1.4_delta} and for a star of mass $2.3~M_\odot$ in fig. \ref{fig:2.3_delta}. 


With the existing pairing interactions the critical temperature of neutron superfluidity ($T_{cn}$) inside the core of NS is in the range $10^{8-9}$ K and the critical temperature of proton superconductivity ($T_{cp}$) is above $10^9$ K for all permissible models which vary with the density of matter inside the core as well. 

\begin{figure}
  \centering
    \includegraphics[width=\linewidth]{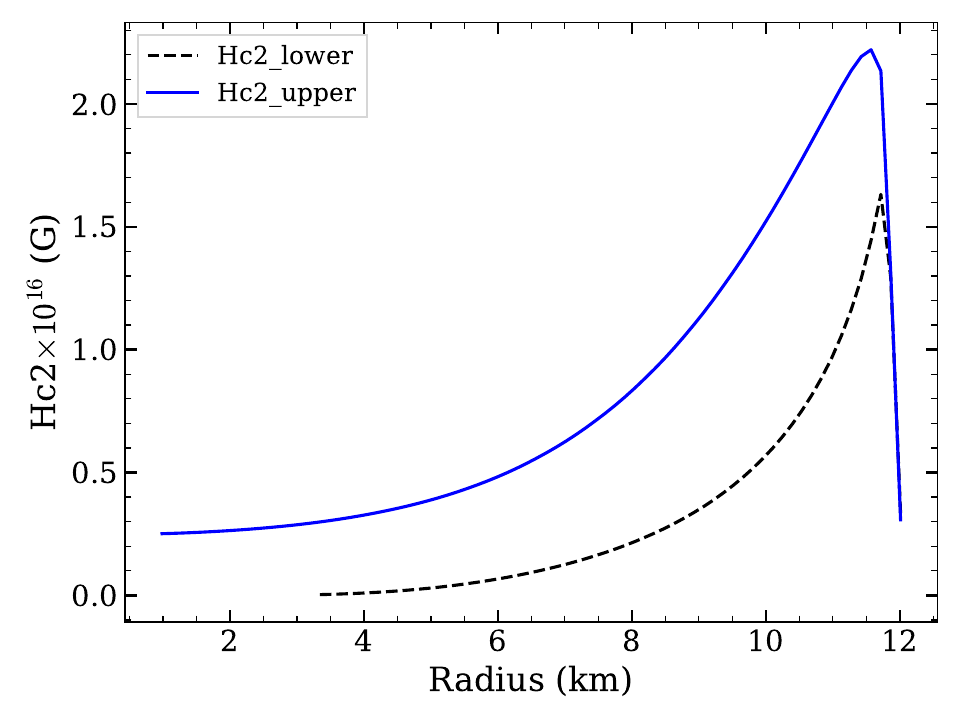}
    \caption{\yn{Upper critical field (dotted) and lower critical field for 1.4$M_\odot$ star.}}
    \label{fig:1.4_critical_field}
\end{figure}

\begin{figure}[t]
    \includegraphics[width=\linewidth]{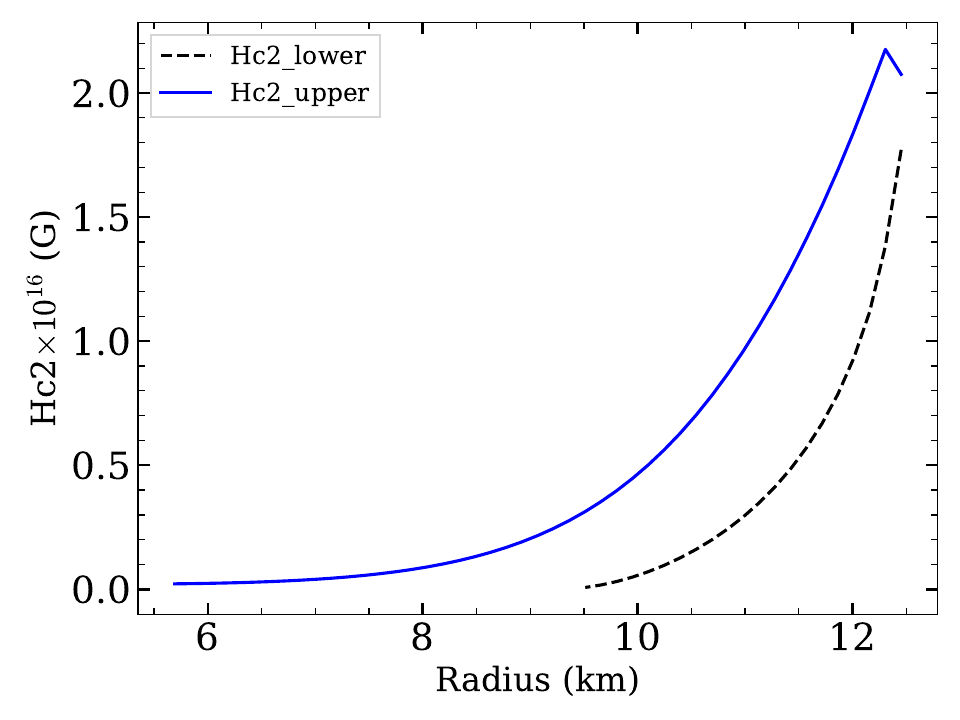}
    \caption{\yn{Upper critical field (dotted) and lower critical field for 2.3$M_\odot$ star.}}
    \label{fig:2.3_critical_field}
\end{figure}

On the other hand, proton superconductivity also depends on the present magnetic field strength. If the present magnetic field is higher than the upper critical field $H_{c2}$ then the proton pairing will be destroyed. Again with the existing models of proton pairing inside the stars $H_{c2}$ is much higher near the surface compared to that at the inner core region.  Hence, depending on the considered models, neutron superfluidity and proton superconductivity will be present inside the entire core or in part of the core. Correspondingly, we also studied the upper critical field $H_{c2}$ with radius. Their plot has been given in fig. \ref{fig:1.4_critical_field} and fig. \ref{fig:2.3_critical_field}. By taking all those into account, we find three different regions inside the NSs. Then, we compute the ambipolar velocity profile inside the star accordingly.


\section{Evaluation of Ambipolar velocity profile}
Now, with this model stars, we solve for the profile of $V_I$ within the star solving eq. \eqref{strsumpedeffnaccsfgen} numerically \ms{for different regions. To solve the equation we note some valid approximations can be considered.} 

\paragraph{Approximation 1}
First we note that the interaction forces arise from particle collision can be approximated by neglecting ${D}_{ne}$ \citep{2011MNRAS.413.2021G} as neutron-proton scattering (mediated by strong interaction) is much more frequent than neutron-electron scattering (mediated by weak interaction). \ms{Then eq. \eqref{ambiii} can be approximated as}

\begin{equation}
   \mathcal{R}_{np}D_{np} \mathbf{V}_p \approx D_I\mathbf{V}_I ~~~ {\rm and} ~~~ {\cal R}_{np}D_{np} = D_I
    \label{eq:DI}
\end{equation}
Then we can write \yn{from Eqs. \eqref{eq:vpve},\eqref{eq:ab} and \eqref{eq:DI} }
\begin{equation}
     \mathbf{V_p} = \mathbf{V_I} ~~~ {\rm and} ~~~ \mathbf{V_e} = \mathbf{V_I} - \frac{\mathbf{j}}{en_c}
\end{equation}
\mg{with the approximations
\begin{equation}
    A = 0, ~~~ {D_I} = {\cal R}_{np} D_{np}, ~~~ {\rm and} ~~~ B = \frac1{en_c}.
\end{equation}}

Collision frequency is given by \citep{1991Ap&SS.176..191Y}
\begin{equation}
    D_{np} = \yn{\rho_p}\,6.6\times10^{16}T_8^2\rho_{14}^{-\frac{1}{3}} \quad s^{-1}
\end{equation}
and the reduction factor $\mathcal{R}_{np}$ \citep{2011MNRAS.413.2021G}
\begin{equation}
    \mathcal{R}_{np} \approx 0.01y_n^2y_pe^{-(y_n+y_p)}
\end{equation}
where $y_n = 1.188\,T_{cn}/T$ and $y_p = 1.764\,T_{cp}/T$\\

\ms{For superfluid nucleonic matter we have the following approximations.}
\paragraph{Approximation 2}
In our model, the neutron vortex density is \citep{tilleyandtilley}
\begin{align}
    \begin{split}
        n^V_n \sim \frac{4\Omega m_N}{h} \sim 10^4\quad cm^{-2}
    \end{split}
\end{align}
for a rotation period of 10s. 

On the other hand, 
proton vortex density for magnetic field $\sim~10^{15}$G  is \citep{2022sai..book.....C}
\begin{align}
    \begin{split}
        n^V_p \sim \frac{2eB_0}{h} \sim 4.83\times 10^{20}\quad cm^{-2}
    \end{split}
\end{align}
which is way more larger than the neutron vortex density. This allows us to safely neglect vortex force due to neutron $\mathbf{f}_n^v$ in eqs. \eqref{eq:final1}.

\paragraph{Approximation 3}
The magnitude for the London field is given by  \citep{tilleyandtilley}
\begin{align}
\begin{split}
    B_L \sim \frac{2m}{e}\Omega
    \sim 10^{-4} G
    \label{londonfield}
\end{split}
\end{align}
for rotation period 10s which is $10^{19}$ times smaller than the external magnetic field. So, we can safely neglect it by comparing  with the present external magnetic field.

\mg{Now, $\mathbf{f}^v_p = \mathbf{f}^m - \mathbf{f}^L$, this will cancel other $\mathbf{f}^L$ term in eqs. \eqref{strsumpedeffnaccsfgen}, where 
\begin{equation}
    \mathbf{f}^m = \frac{m_n}{\mu_0 e}\left[ - \mathbf{\bar \omega_p} \times \left\{\nabla\times(H_{c1}\hat{\bar \omega}_p) \right\} + \frac{1}{m} \left(\vec{\nabla}\times\vec{p^p}\right) \times \left(\vec{\nabla} \times \vec{B_L}\right) \right] 
    \label{magforce}
\end{equation}
From eqs. \eqref{londonfield} it is clearly visible that London field is too weak compare to magnetic field of magnetar. So we can neglect the London field term in eqs. \eqref{magforce}. Also it can be shown $H_{vp} = H_{c1}$ \citep{2011MNRAS.410..805G}. So $\mathbf{f}^m$ becomes 
\begin{equation}
    \mathbf{f}^m = \frac{m_N}{\mu_0 e}\left[ - \mathbf{\bar \omega_p} \times \left\{\nabla\times(H_{c1}\hat{\bar \omega}_p) \right\}\right]
    \label{finalmagforce}
\end{equation}}

 These considerations allow further simplification of eq. \eqref{eq:final1} \ms{for superfluid nucleonic matter, which reduces to}.

\begin{align}
    \begin{split}
        \frac{\mathbf{f}^m}{\rho}&=x_p \mathbf{\nabla}\beta + 2x_p\frac{D_I}{\rho_n}\mathbf{V}_I - 2ax_p\mathbf{\omega_n} \times \mathbf{V}_I\\
        &- 2x_p\hat{\omega}_n \times \left[ \mathbf{\omega}_n \times \left\{ b\left(\mathbf{V}_I - \frac{\mathbf{j}}{en_c}\right)\right\}\right]+2x_p\left(\mathbf{\Omega} \times \mathbf{V}_I\right).
        \label{finale}
    \end{split}
\end{align}

\ms{In the above eq. \eqref{finale} the} \mg{first term is transfusion which is related to weak interaction maintaining the $\beta$-equilibrium.}

Now, \yn{from Eq. \eqref{transfusion}}
\ms{\begin{align}
    \begin{split}
    \beta & = - \frac{n_c}{\lambda}\mathbf{\nabla}\cdot\mathbf{w}\\
    & = - \frac{n_c}{\lambda}\mathbf{\nabla}\cdot\left(R_{np}\mathbf{V}_I\right)\\
    & = -\frac{n_c}{\lambda}\left[ \frac{1}{r}\frac{\partial}{\partial r}\left(R_{np}r\mathbf{V}_{Ir}\right)+\frac{\partial}{\partial z}\left(R_{np}V_{Iz}\right)\right].
    \end{split}
\end{align}} 
\ms{The term for differentiation with respect to $\phi$ is absent because of the rotational symmetry of the rotating star. And
\begin{equation}
    \mathbf{w}
    = {\cal R}_{np} \mathbf{V_I} + \frac1{2en_c} \mathbf{j},
\end{equation}
with approximation 1.}

\ms{Then we considering cylindrical coordinate system and star rotation along $z$-axis we have given the component wise equations in \ref{sec:App:C}. }

\mg{We use the $\lambda$ value from \cite{2011MNRAS.413.2021G} for mUrca, 
\begin{equation}
    \lambda = 2.5\times 10^{40}\rho_{14}^{\frac{2}{3}}T_8^6 J^{-1}m^{-3}s^{-1}
\end{equation}
$n_c$ can be obtained from EoS.}

Unlike this superfluid nucleonic matter, the only neutron superfluid matter and normal matter equations do not have such an approximation. However, there will be the absence of a few terms from the equation \eqref{eq:final1}. So the ambipolar velocity equation governing only neutron superfluid matter will be from eq. \eqref{eq:final2}.
Similarly, equation for completely normal matter will be from eq. \eqref{eq:normal_matter}.
$a$ and $b$ can be calculated from \cite{2011MNRAS.413.2021G}.

From the figures (\ref{fig:1.4_delta}),  (\ref{fig:2.3_delta}), (\ref{fig:1.4_critical_field}) and (\ref{fig:2.3_critical_field}), it is apparent that various regions inside the core will arise. For $1.4~M_\odot$ star, there will be two regions-- \textbf{a)} only neutrons are superfluid, and both proton, electrons are normal \textbf{b)} both neutrons and protons are superfluid and only electrons are normal for both upper and lower energy gap. The same thing is for $2.3~M_\odot$ except there is an extra region where all the constituent particles are in normal phase.   

\begin{figure}
  \centering
    \includegraphics[width=\linewidth]{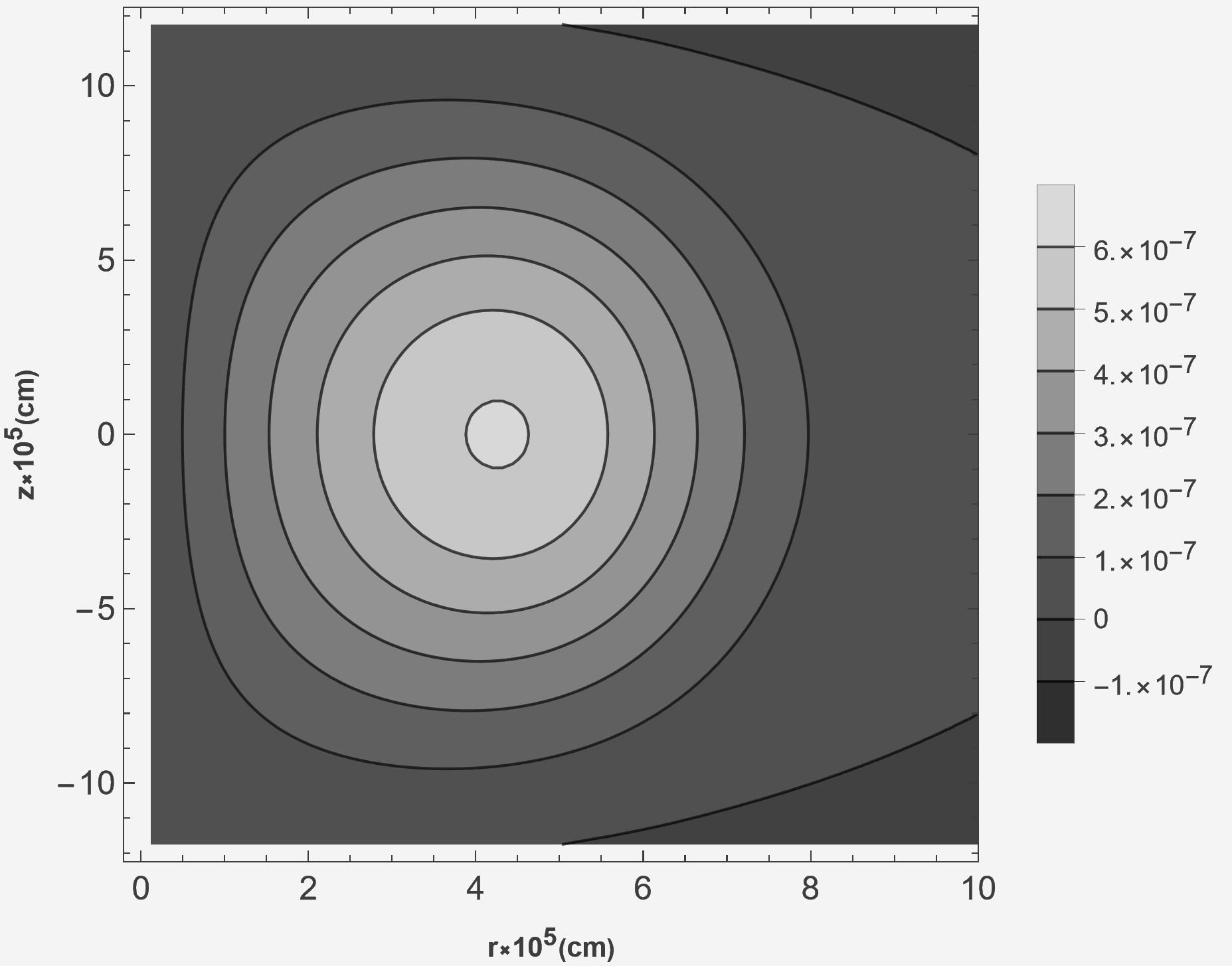}
    \caption{Radial component of ambipolar velocity for $1.4~M_\odot$}
    \label{fig:1.4_velocity_uvr}
\end{figure}

\begin{figure}[t]
    \includegraphics[width=\linewidth]{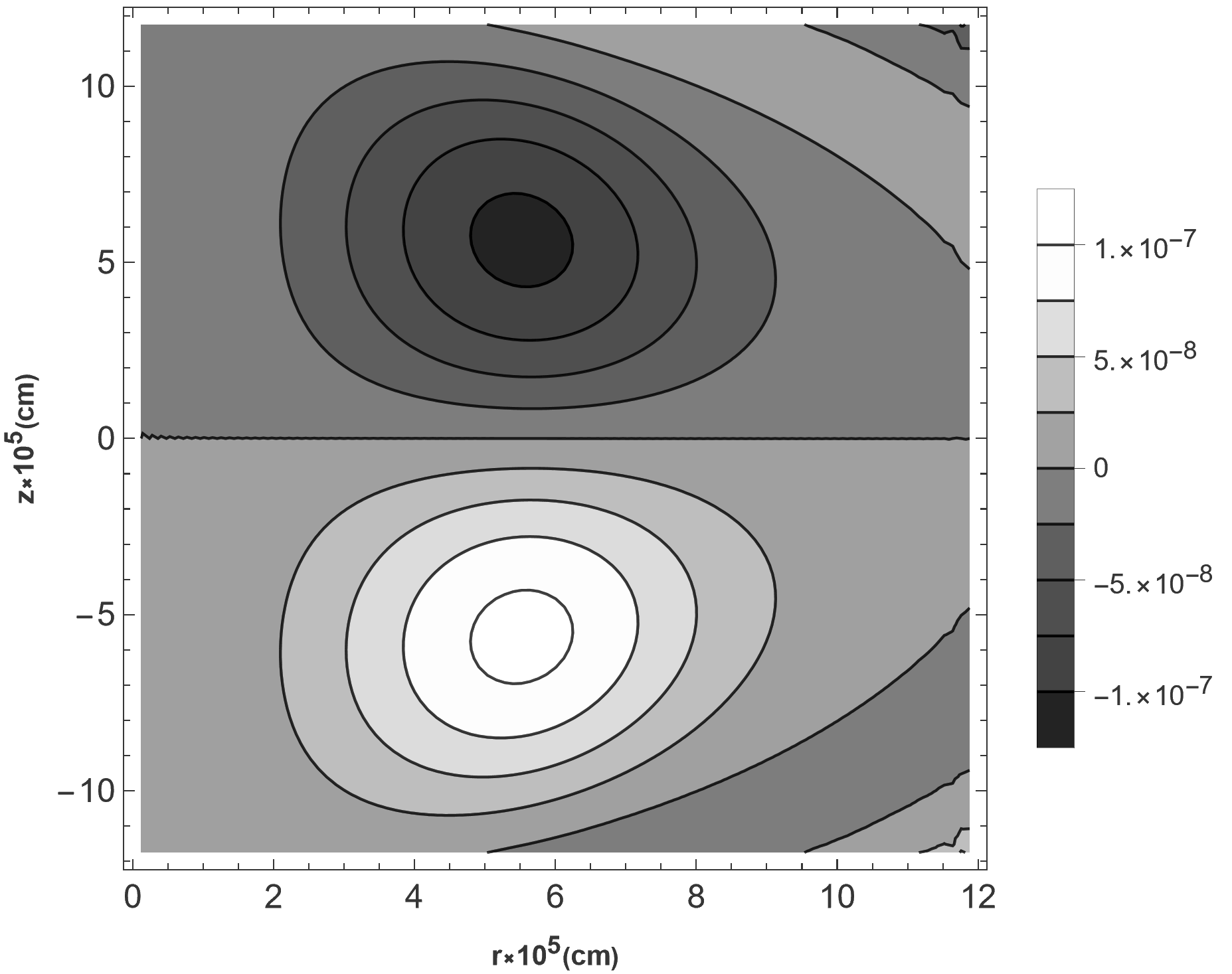}
    \caption{Axial component of ambipolar velocity for $1.4~M_\odot$}
    \label{fig:1.4_velocity_uvz}
\end{figure}

Finally, we compute the eq. \eqref{finale} for two different stars with masses 1.4 and 2.3 $M_\odot$ respectively in different regions. For the first time, we present the detailed profile of radial and axial components of ambipolar velocity inside the core of the magnetar. Figure (\ref{fig:1.4_velocity_uvr})  and (\ref{fig:1.4_velocity_uvz}) represent ambipolar velocity profile in the core of $1.4~M_\odot$ star consider different pairing gaps of proton and neutron. 
 
The azimuthal component has the same profile (a consequence
of the poloidal nature of magnetic field) as the radial component, its magnitude being negligible.

Now, using \eqref{rate_normal_magnetic_enr_decay} and \eqref{rate_super_magnetic_ener_decay} we computed energy decay rate, 
\begin{equation}
\left(\frac{dE_b}{dt}\right)_{amb} = - 9.68\times10^{35}\quad erg/s
\end{equation}
\begin{equation}
\left(\frac{dE_b}{dt}\right)_{amb} = - 1.12\times10^{36}\quad erg/s
\end{equation}

for $1.4~M_\odot$ and $2.3~M_\odot$ star respectively. 

The relevant time scale for magnetic field evolution inside magnetar core is \mg{given by}
\mg{
\begin{align}
    t = \frac{L}{V},
\end{align}
where $L$ is the characteristic length over which magnetic field varies and $V$ is the
average of the magnitude of $V_I$. This gives} $t \approx 6.72\times10^4$ years and $t \approx 3.17\times10^4$ years for $1.4~M_\odot$ and $2.3~M_\odot$ star respectively. \mg{This time scale for magnetic field evolution indicate the age of the magnetar. Combining this with energy decay rate would place these magnetars in the similar positions as the observed ones (see Fig. \ref{fig:magnetar_plot}).} \yn{The figure demonstrates that the energy decay rate and their evolution with characteristic age are in close agreement with the observed values over the relevant range.} Subsequently, we also calculated the magnetic field decay with time by eq. \eqref{field_evolution}. The computation yields an average value of 0.06 $Gs^{-1}$ and 0.09 $Gs^{-1}$ for $1.4~M_\odot$ and $2.3~M_\odot$ star respectively.

\section{Discussion}\label{sec:dis}
The magnetic field associated with the NSs decays with
time by several possible mechanisms depending on the composition
and the surroundings of the constituent matter. In the core 
of the star, the field mainly evolves due to the ambipolar diffusion
of the charged particles. The magnetic force on the charged 
particles is balanced by pressure gradient force and inter-particle
interaction. We present here the charged particle dynamics in 
the background of neutrons. As the magnetic force is balanced
by both transfusion and inter-particle interaction we have discussed 
the different possibilities of particle interaction depending 
on the state of the matter - whether it is in the normal phase or 
in the superfluid state. We take a general expression of particle 
interaction which can take care of all possibilities, i.e. all 
particles are in the superfluid state, some of the species are in 
superfluid state, or all the species are in the normal phase. We have
also considered the force of vortices in the case of superfluid matter
which is zero for normal species. With 
this general procedure, we arrive at a general equation for ambipolar 
velocity. By solving the equation with different scenarios of the 
core as it evolves with temperature and time, the overall picture 
of the magnetothermal evolution of the core will be clear. For 
example, immediately after the birth of the star, the core temperature 
is considerably high. Consequently, that time it is reasonable 
to consider the matter in its normal state and the inter-particle 
interaction is dominating contribution to balancing the magnetic 
force. According to that the general ambipolar velocity equation 
can be considered with some term absent in the equation (eq. 
(\ref{strsumpedeffnaccsfgen})). Henceforth it can be solved 
for that particular age of the star. Subsequently, with the 
decrease in the temperature of the core, the probability of 
neutron and proton pairing emerges leading to nucleon superfluidity. 
Again the transition from normal phase to superfluid phase will 
occur separately for neutrons and protons at different temperatures 
depending upon the critical temperature $T_c$  of superfluidity 
for neutrons and protons. So, there may be a phase, when one of 
the nucleons is in normal phase and the other in superfluid 
phase. Then also there will arise two special cases with different 
specifications of the superfluid inter-particle interaction. 
At the same time the relative strength of transfusion and inter-particle 
interaction will alter. With the specific case of normal matter, employing this general model and considering 
the entire star is made of normal nucleons and electrons, at 
temperature $\sim 10^8$ K, the typical timescale of ambipolar diffusion
comes out to be $\sim 9\times 10^4$ yr \citep{2021MPLA...3650144B}, which is very 
close to the age of magnetars. Other estimates show either 
very short \citep{2018PhRvD..98d3007O,2020MNRAS.498.3000C} or 
very long timescale \citep{2011ApJ...740L..35G,2017MNRAS.471..507C} 
of field evolution compared to the typical age of magnetars.
 In our subsequent work, we shall discuss in detail ambipolar velocity and field evolution, by invoking superfluid nucleons (under preparation).
Hence, switching on and off different terms in eq. (\ref{strsumpedeffnaccsfgen})
it is possible to get the ambipolar velocity profile inside 
the core of a star at different temperature regimes. The ambipolar
velocity profile is pivotal to calculate the magnetic
field decay and hence the field energy decay. This will in turn 
help to get an idea of the magnetar activities during the quiescent 
period. At different ages of the star, the internal heating 
of the core due to field decay can also be calculated following
this calculation which will be useful in cooling simulation. 
Moreover, the procedure can be extended in a very straightforward 
manner to matter with other exotic components such as heavier 
baryon.

\section*{Acknowledgments}
The authors acknowledge the funding support from the Science and Engineering Research Board, the Department of Science and Technology, and the government of India through project no. CRG/2022/000069.

\appendix
\section{Normal matter Electric field}

In this appendix, we work out the explicit expression for the electric field in normal matter. Ambipolar velocity is already defined in eq. \eqref{nambipolar_velocity}
Now, we solve eq.\eqref{nambipolar_velocity} and eq. \eqref{ncurden1} to write electron and proton velocity in terms of ambipolar velocity and current density. Then, we solve eq. \eqref{streulpden4} and \eqref{streuleden4} and finally obtain our desired electric field equation,

\begin{align}
\begin{split}
        \mathbf{E} = \left[\frac{1}{\frac{1}{D_{pn}} + \frac{1}{D_{en}}} + D_{pe}\right]\frac{\mathbf{j}}{n_c^2e^2} + \frac{D_{pn} - D_{en}}{D_{pn} + D_{en}}\frac{\mathbf{j}\times\mathbf{B}}{n_ce}
        \nonumber\\
        - \left(\mathbf{V}_I \times \mathbf{B}\right) + \frac{\frac{\mathbf{\nabla}\mu_p}{D_{pn}} - \frac{\mathbf{\nabla}\mu_e}{D_{en}}}{e\left(\frac{1}{D_{pn}} + \frac{1}{D_{en}}\right)}.
\end{split}
\end{align}

Collision force $\mathbf{f}^c_n$ have been discussed in subsection (\ref{sec: normal_matter}) and neglecting inertia term. 
\section{Superfluid matter electric field}
In this appendix, we obtain the expression for electric when nucleons are in superfluid state. We have derived this electric field expression just like the previous one except our ambipolar velocity is defined in eq. \eqref{superfluid_ambipolar}. Different interaction forces have been discussed in section (\ref{sec: superfluid_nucleonic_matter}). 

\begin{align}
\begin{split}
       \mathbf{E} &= \left[\frac{1}{\frac{1}{\mathcal{R}_{pn}D_{pn}} + \frac{1}{\mathcal{R}_{en}D_{en}}} + \mathcal{R}_{pe}D_{pe}\right]\frac{\mathbf{j}}{n_c^2e^2}
        + \frac{\frac{\mathbf{\nabla}\mu_p}{\mathcal{R}_{pn}D_{pn}} - \frac{\mathbf{\nabla}\mu_e}{\mathcal{R}_{en}D_{en}}}{e\left(\frac{1}{\mathcal{R}_{pn}D_{pn}} + \frac{1}{\mathcal{R}_{en}D_{en}}\right)}\\
        &- \left[1 - \frac{a_{pn}\rho_p}{m_N\left(1 + \frac{\mathcal{R}_{pn}D_{pn}}{\mathcal{R}_{en}D_{en}}\right)}\right](\mathbf{V}_I\times\mathbf{B})\\
        &+ \left[\frac{\mathcal{R}_{pn}D_{pn} - \mathcal{R}_{en}D_{en}}{\mathcal{R}_{pn}D_{pn} + \mathcal{R}_{en}D_{en}} + \frac{a_{pn}\rho_p}{m_N}\frac{1}{\left(1 + \frac{\mathcal{R}_{pn}D_{pn}}{\mathcal{R}_{en}D_{en}}\right)^2}\right]\frac{\mathbf{j}\times\mathbf{B}}{n_ce}\\
        &- \frac{1}{n_ce\left(1 + \frac{\mathcal{R}_{pn}D_{pn}}{\mathcal{R}_{en}D_{en}}\right)}\left[a_{pn}\mathbf{\omega}_p\times\rho_p\mathbf{V}_p + a_{pn}\mathbf{\omega}_p \times (\mathbf{\nabla} \times \lambda_p\hat{\omega}_p)\right]\\
        &- \frac{1}{n_ce\left(1 + \frac{\mathcal{R}_{pn}D_{pn}}{\mathcal{R}_{en}D_{en}}\right)}\left[\frac{a_{pn}e}{m_N}\mathbf{B}\times (\mathbf{\nabla}\times \lambda_p\hat{\omega}_p) + \mathbf{F}^v_p\right]\\
        &- \frac{1}{n_ce}\mathbf{F}^{sf}_{pe} - \frac{1}{n_ce\left(1 + \frac{\mathcal{R}_{en}D_{en}}{\mathcal{R}_{pn}D_{pn}}\right)} \left[b_{ne}\hat{\omega}_n\times \left\{\mathbf{\omega}_n \times \rho_n\mathbf{V}_e\right\}\right]\\ 
       &+ \frac{1}{n_ce\left(1 + \frac{\mathcal{R}_{en}D_{en}}{\mathcal{R}_{pn}D_{pn}}\right)}\left[b_{ne}\hat{\omega}_n\left\{\mathbf{\omega}_n \times \mathbf{\nabla}\times\lambda_n\hat{\omega}_n\right\}\right]
\end{split}
\end{align}

where $\mathbf{F}^{sf}_{pe}$ is interaction between proton vortex with electron.

Now, when proton loses its superconducting state and becomes normal matter, it left us with only neutron superfluid matter. In this state electric field doesn't change drastically, it's just a matter of fewer terms as there is no vortex-vortex interaction. This left us with the expression
\begin{align}
\begin{split}
   \mathbf{E} &= \left[\frac{1}{\frac{1}{\mathcal{R}_{pn}D_{pn}} + \frac{1}{\mathcal{R}_{en}D_{en}}} + D_{pe}\right]\frac{\mathbf{j}}{n_c^2e^2} - \left(\mathbf{V}_I \times \mathbf{B}\right)\\
   &+ \left[\frac{\mathcal{R}_{pn}D_{pn} - \mathcal{R}_{en}D_{en}}{\mathcal{R}_{pn}D_{pn} + \mathcal{R}_{en}D_{en}}\right]\frac{\mathbf{j}\times\mathbf{B}}{n_ce} + \frac{\frac{\mathbf{\nabla}\mu_p}{\mathcal{R}_{pn}D_{pn}} - \frac{\mathbf{\nabla}\mu_e}{\mathcal{R}_{en}D_{en}}}{e\left(\frac{1}{\mathcal{R}_{pn}D_{pn}} + \frac{1}{\mathcal{R}_{en}D_{en}}\right)}\\
   &+ \frac{1}{n_ce\left(\frac{1}{\mathcal{R}_{pn}D_{pn}} + \frac{1}{\mathcal{R}_{en}D_{en}}\right)}\left(\frac{\mathbf{f}^{sf}_{pn}}{\mathcal{R}_{pn}D_{pn}} - \frac{\mathbf{f}^{sf}_{en}}{\mathcal{R}_{en}D_{en}}\right).
\end{split}
\end{align}

$\mathbf{f}^{sf}_{pn}$ and $\mathbf{f}^{sf}_{en}$ are the interaction forces discussed at the end of the section (\ref{sec: superfluid_nucleonic_matter}).

\mg{\section{Differential Equations in component wise form}
\label{sec:App:C}
The component wise the first term in eq. \eqref{finale} $  \frac{\partial \beta}{\partial r}$ and $\frac{\partial \beta}{\partial z}$ are given by
\begin{align}
    \begin{split}
    \frac{\partial \beta}{\partial r} & = - \frac{\partial}{\partial r}\left(\frac{n_c}{\lambda}\right)\left[ \frac{1}{r}\frac{\partial}{\partial r}\left(rR_{np}V_{Ir}\right) + \frac{\partial}{\partial z}(R_{np}V_{Iz})\right] \\
    & - \frac{n_c}{\lambda}\left[ \frac{1}{r}\frac{\partial^2}{\partial r^2}\left(rR_{np}V_{Ir}\right) - \frac{1}{r^2}\frac{\partial}{\partial r}\left(rR_{np}V_{Ir}\right) + \frac{\partial}{\partial z}(R_{np}V_{Iz})\right] \label{finalbetar}
    \end{split}
\end{align}
And
\begin{align}
    \begin{split}
    \frac{\partial\beta}{\partial z} = - \frac{n_c}{\lambda}\left[ \frac{1}{r}\frac{\partial^2}{\partial r \partial z}\left(rR_{np}V_{Ir}\right) + \frac{\partial^2}{\partial z^2}(R_{np}V_{Iz})\right] \label{finalbetaz}
    \end{split}
\end{align}
Taking the rotation along z-axis in cylindrical coordinate and putting all values together we will get three equation 
\begin{align}
        x_p\frac{\partial \beta}{\partial r} + 2x_p\left(\frac{D_I}{\rho_n} + \omega_n b_{ne}\right)V_{Ir} + 2x_p\left(\omega_n a - \Omega\right)V_{I\phi} = \frac{f^m_r}{\rho}\\
        x_p\frac{\partial \beta}{\beta z} + \frac{D_I}{\rho_n}V_{Iz} = \frac{f^m_r}{\rho}\\
        And\quad V_{I\phi} = \frac{\left(\omega_n a - \Omega\right)V_{Ir} + \omega_n b_{ne}Bj}{\frac{D_I}{\rho_n} - \omega_n \beta_{ne}}
\end{align}
Now substituting values of $V_{I\phi}$\quad and\quad $\frac{\partial \beta}{\partial r}$\quad $\frac{\partial \beta}{\partial z}$ in above we finally obtain
\begin{align}
    \begin{split}
        & - x_p\frac{\partial}{\partial r}\left(\frac{n_c}{\lambda}\right)\left[ \frac{1}{r}\frac{\partial}{\partial r}\left(rR_{np}V_{Ir}\right) + \frac{\partial}{\partial z}(R_{np}V_{Iz})\right] \\
    & - \frac{x_p n_c}{\lambda}\left[ \frac{1}{r}\frac{\partial^2}{\partial r^2}\left(rR_{np}V_{Ir}\right) - \frac{1}{r^2}\frac{\partial}{\partial r}\left(rR_{np}V_{Ir}\right) + \frac{\partial^2}{\partial z\partial r}(R_{np}V_{Iz})\right] \\
    & + 2x_p\left[\frac{D_I}{\rho_n} + \omega_n b_{ne} + \frac{\left(\omega_n a - \Omega\right)^2}{\frac{D_I}{\rho_n} - \omega_n b_{ne}} \right]V_{Ir} \\
    &+ \frac{2x_p b_{ne}\omega_nB\left(\omega_n a - \Omega\right)}{\frac{D_I}{\rho_n} - \omega_n b_{ne}}j = \frac{f^m_r}{\rho}
    \end{split}
\end{align}
And
\begin{align}
    \begin{split}
        - x_p\frac{n_c}{\lambda}\left[ \frac{1}{r}\frac{\partial^2}{\partial r \partial z}\left(rR_{np}V_{Ir}\right) + \frac{\partial^2}{\partial z^2}(R_{np}V_{Iz})\right] + \frac{2x_p}{\rho_n}D_IV_{Iz} = \frac{f^m_z}{\rho}.
    \end{split}
\end{align}}
  
\mg{To close the system of equations, we enforce a boundary condition where the ambipolar velocity identically vanishes at the boundaries. These coupled equations are solved subject to the boundary condition that $\mathbf{V}_I$ vanishes at the boundaries. }
\bibliographystyle{elsarticle-num-names} 
\bibliography{pla_ambipolar}

\end{document}